\documentclass[a4paper,12pt]{article}
\usepackage{graphicx}
\usepackage[top=3.0cm,bottom=3.0cm,left=3.0cm,right=2.0cm]{geometry}
\usepackage{amssymb,amsmath}
\usepackage{textcomp}
\usepackage{subfigure}
\usepackage{cite}
\usepackage{braket}
\usepackage{xcolor}

\newcommand{\QCD}{{\textrm{\scriptsize QCD}}}

\title{ Fractals, non-extensive statistics and QCD}

\author{Airton Deppman$^{1,2}$, Eugenio Meg\'ias$^{2}$, Debora P. Menezes$^{3}$}
\date{1- Instituto de F\'isica -  Universidade de S\~ao Paulo \\ email: deppman@if.usp.br ; \\  2- Departamento de F{\'\i}sica At\'omica, Molecular y Nuclear and
  Instituto Carlos I de F{\'\i}sica Te\'orica y Computacional \\ Universidad
  de Granada, E-18071 Granada, Spain;\\  3- Universidade Federal de Santa Catarina.}

\begin{document}

\maketitle

\begin{abstract}
 In this work we analyse how scaling properties of Yang-Mills field theory manifest as self-similarity of truncated n-point functions by scale evolution.
 The presence of such structures, which actually behaves as fractals, allow for recurrent non-perturbative calculation of any vertex. Some general properties are indeed independent of the perturbative order, what simplifies the non-perturbative calculations. We show that for sufficiently high perturbative orders a statistical approach can be used, the non extensive statistics is obtained, and the Tsallis index, $q$, is deduced in terms of the field theory parameters. The results are applied to QCD in the one-loop approximation, where $q$ can be calculated, resulting in a good agreement with the value obtained experimentally. We discuss how this approach allows to understand some intriguing experimental findings in high energy collisions, as the behavior of multiplicity against collision energy, long-tail distributions and the fractal dimension observed in intermittency analysis.
\end{abstract}


\section{Introduction}

Yang-Mills field (YMF) theory is a prototype theory to describe three among the four known natural interactions, namely, strong, weak and electromagnetic, with only gravitational force left aside. The importance of such cathegory of field theory is summarized in the Satandard Model, which is a single framework to describe all three interactions in a unified formalism. Scale invariance is a fundamental aspect of the Yang-Mills field theory, playing an important role in the renormalization of the theory after the ultraviolet divergences are subtracted~\cite{Dyson1, Ward, Gell-Mann_Low}.  The theory was extremely succesful in QED and Weak interactions, providing accurate and precise descriptions for the observed phenomena. QCD has proved to be more challenging, since the calculational methods applied to the other two forces is not appropriate to provide accurate results for strong interacting systems. In this scenario. QCD has been tested in high energy elementary particle interactions, where asymptotic freedom allows for the application of perturbative methods, and by Lattice QCD (LQCD), which is a numerical approach based on QCD first principles.

Fractals are complex systems with internal structure presenting scale invariance and self-similarity. Fractal measures, contrary to more conventional quantities for which an increase in resolution results in the same measured value with increased precision, yield different values for different resolutions. A classical example is the length of coastlines~\cite{Mandelbrot}. Since its proposal about three decades ago, the concept of fractals has found many applications in Mathematics, Arts, Biology and Complex Systems in general. A nice introduction to the subject and its applications can be found in Refs.~\cite{Mandelbrot,West}, and more formal descriptions in Ref.~\cite{Falconer}. Among the most important features of fractals are the  scaling properties, where the internal structure of the fractal is equal to the main fractal, but with a reduced scale, self-symmetry and fractional dimensions.

Tsallis statistics generalizes Boltzmann-Gibbs one by introducing a non-additive form for Entropy, what leads to non extensivity of some quantities. The effects of Tsallis statistics have been explored since it was proposed, in the late 80's   \cite{Tsallis1988,TsallisBook}, but its full meaning and fundamentals are still far from being completely understood. The $q$-deformed entropy functional that underlies non extensive statistics depends on a real parameter, $q$,
that determines the degree of non additivity of the functional and in
the limit $q \rightarrow 1$, it becomes additive and the standard
Boltzmann-Gibbs entropy is recovered. 

Our goal is to show a subtle link between scale invariance of YMF, fractals and the Tsallis non extensive statistics. For that aim we analyse
how the scaling properties of a Yang-Mills field theory leads to recurrence relations that allow non-perturbative calculations, which amounts to a self-similar behaviour of  truncated  n-point  diagrams by scale evolution.  Essentially, these amplitudes are shown to behave as fractals by evolving the scale to the ultraviolet region and the calculation of vertexes even in high perturbative orders  becomes possible due to a simple recurrence formula. For sufficiently high perturbative order the non extensive statistics is obtained, and the Tsallis index, $q$, is deduced in terms of the field parameters. 

Our work is organised as follows: we first introduce the basic concepts of the Yang Mills theory and how a sistem of interacting partons relates to fractals. Then a connection between the partonic self-similarity and the non-extensive statistics is established.  In other words, we will demonstrate  that renormalizable field theories lead to fractal structures, which can be studied, from a thermodynamical point of view, with Tsallis statistics.
Finally, as an application, well known QCD scalling properties are used to calculate, for the first time, the $q$ value in terms of QCD fundamental parameters. Some consquences of our resuts are discussed, in particular the fractal dimension, which is experimentally observed by intermittency analysis, and the behaviour of particle multiplicity as a function of the collision energy, which is here related to the non extensivity and to the fractal dimension of the hot and dense system formed at high energy collisions. This is an intriguing aspect of experimental data which can be explained in a simple way from QCD by the present approach, and some proprties of high energy collisions, such as multiplicity dependence on the collision energy, long-tail distributions and fractal dimension observed in intermittency studies, are described in our approach.

\section{Formalism}

The simplest scale free non-abelian gauge field theory has Lagrangian density including bosons and fermions given by
\begin{equation}
 {\cal L}=-\frac{1}{4} F^{a}_{\mu \nu} F^{a \mu \nu}+ i \bar{\psi_j} \gamma_{\mu} D^{\mu}_{ij} \Psi_j \label{simplestgauge}
\end{equation}
where
\begin{equation}
  F^{a}_{\mu \nu}=\partial_{\mu} A^{a}_{\nu} - \partial_{\nu} A^{a}_{\mu}+g f^{abc}A^{b}_{\mu}A^{c}_{\nu}
\end{equation}
and
\begin{equation}
  D^{\mu}_{ij}= \partial_{\mu}\delta_{ij}-igA^{a \, \mu} T^{a}_{ij}\,,
\end{equation}
where $\psi$ and $A$ represent, respectively, the fermion and the vector fields, with $f^{abc}$ being the structure constants of the group and $T^{a}$ the matrices of the group generators in the fermion representation.

The UV regularized  vertex functions are related to the renormalized vertex functions with renormalized parameters, $\bar{m}$ and $\bar{g}$, as 
\begin{equation}
  \Gamma(p,m,g)=\lambda^{-D} \Gamma(p,\bar{m},\bar{g}) \,.
\end{equation}
This property is mathematically described by the renormalization group equation, which introduce the beta-function, which allowed to show that QCD is asymptotically free~\cite{Politzer1, Politzer2, GrossWilczek1, GrossWilczek2}. Such equation is known as Callan-Symanzik equation~\cite{Callan, Symanzik1, Symanzik2}, and is given by
\begin{equation}
  \left[M\frac{\partial}{\partial M}  + \beta_g \frac{\partial}{\partial \bar{g}} + \gamma \right]\Gamma=0 \label{CallanSymanzik}
\end{equation}
where $M$ is the scale parameter, and the $\beta$-functions are defined as
\begin{equation}
  \beta_{\bar{g}}=M\frac{\partial \bar{g}}{\partial M}\,.
\end{equation}
$D=D_o+d$, with $D_o$ being the natural dimension of the phase-space. In general, $d$ is not an integer, therefore the scaling dimension, $D$, may be fractionary. The parameter $\gamma$ is the anomalous dimension given by a combination of the scaling dimensions of the fields $\psi$ and $A$. 

\begin{figure}[t!]
 \centering
        \includegraphics[scale=0.35]{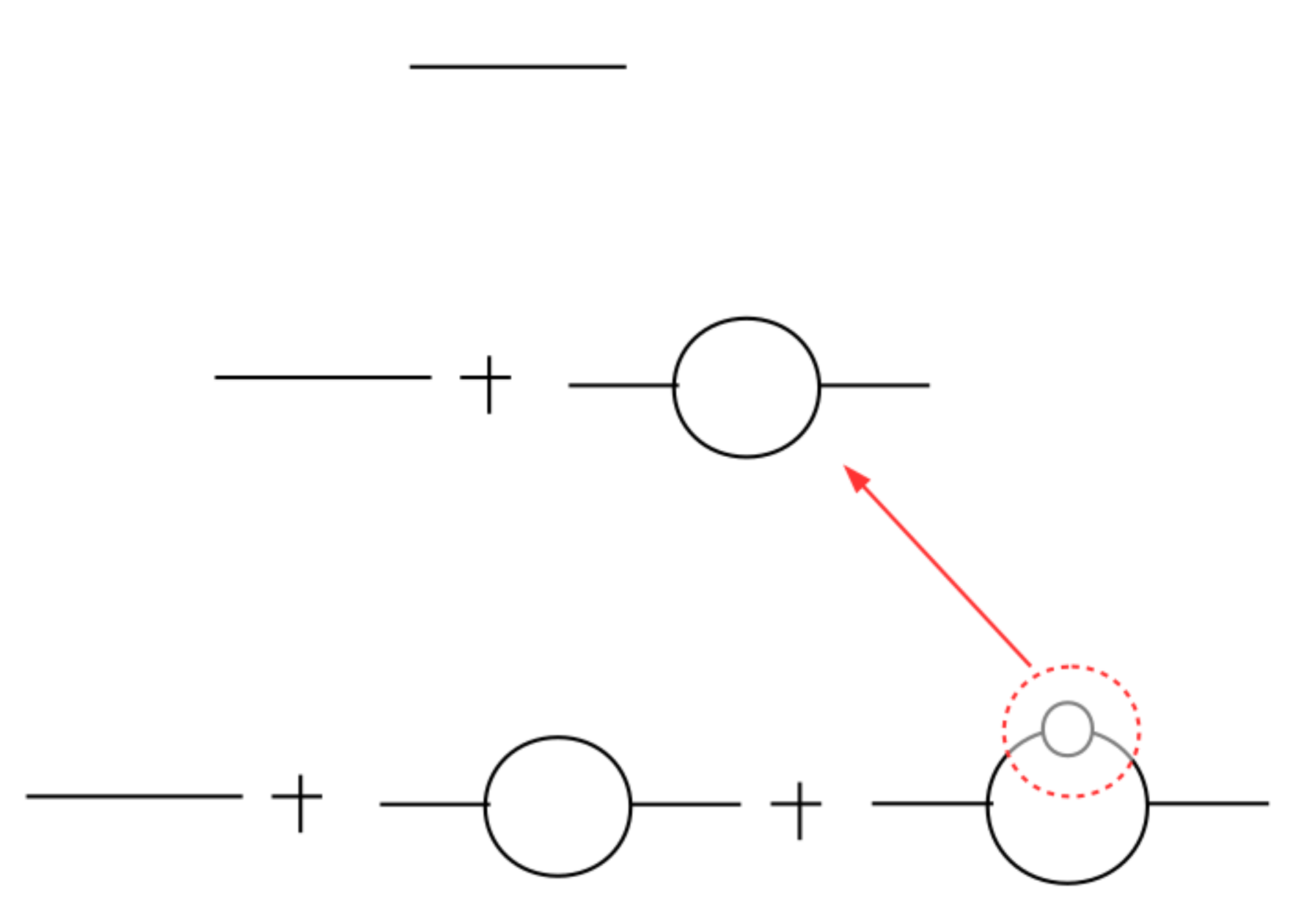} \label{fig:loopscale}
        \caption{Diagrams showing the scaling properties of Yang-Mills fields: a loop in higher order is identical, after proper scaling, to a diagram in lower order.}
\end{figure}

In the irreducible one-particle representation, self-energy is taken into account by using the effective parton mass, $\bar{m}$, what allows a reduction of the calculational complexity, since only proper vertexes must be considered. As schematically shown in Fig.~\ref{fig:loopscale}, 
the scale invariance means that, after proper scaling, the loop in a higher order graph is identical to a loop in lower orders~\cite{Politzer1,Politzer2,GrossWilczek1,GrossWilczek2}. The scale invariance of the vertex function is a direct consequence of the Callan-Symanzik equation, and it is of fundamental importance in what follows. 
When effective charges and masses are used, the line of the respective field in Feynman diagram represent an effective particle or state, since parts of the diagram representing self-energy contribution are omitted because they are already taken into account by the renormalization. Such diagrams are  called irreducible graphs. We will next refer to the physical system represented by irreducible graphs as effective parton. In irreducible graphs, the only allowed vertexes, called proper vertexes, are related to the creation of an effective parton.

Despite the great reduction in complexity when irreducible graphs are used, still complex graphs are unavoidable in performing perturbative calculations, specially for QCD, since vertex functions may include several orders in perturbative approximation. We will show that some additional simplification may be achieved under some circumstances that are relevant in hadron structure and in multiparticle production.
Preliminary results of the approach developed here were discussed in Ref.~\cite{DeppmanMegiasPhysics}.

\subsection{Statistical description of the partonic state}

In this work the quantity of interest is the trace of the evolution operator
\begin{equation}
 Z=\textrm{Tr} \, e^{-iHt} \ket{\Psi_o}\,, \label{Gamma}
\end{equation}
where $\Psi_o$ is an eigenvector of the Hamiltonian $H$ corresponding to the initial state of the fields. $H$ governs the effective parton evolution in the presence of the strong interaction. From now on we refer to proper vertexes as interactions. 

The time evolution of an initial partonic state is
\begin{equation}
 \ket{\Psi} \equiv \ket{\Psi(t)}=e^{-iHt} \ket{\Psi_o} \,.
\end{equation}
The state $\Psi$ can be written as
\begin{equation}
 \ket{\Psi}=\sum_{n} \braket{\Psi_n|\Psi} \ket{\Psi_n}
\end{equation}
with the basis states $\ket{\Psi_n}$ corresponding to a fixed number, $n$, of interactions in the vertex function.  In the perturbative method each proper vertex gives rise to a term in the Dyson series, and at any time, $t$, the partonic state is given by
\begin{equation}
 \ket{\Psi} =  \sum_{\{n\}} (-i)^n \int dt_1 \dots dt_n \bra{\Psi} e^{-iH_o(t_n-t_{n-1})} g  \dots e^{-iH_o (t_1-t_o)}  \ket{\Psi_o},
\end{equation} 
where  $g$ represents the interaction and
$t_n>t_{n-1}>\dots>t_1 >t_o$, and $\sum_{\{n\}}$ runs over all possible terms. 
 with $n$ interaction vertexes. Observe that $\braket{\Psi_{n'}|\Psi_n}=\delta_{n'n}$. The symbol $\{n\}$ indicates that the summation is performed over all possible configuration of fields with $n$ interactions.

The number of particles in the state $\ket{\Psi_n}$ is not directly related to $n$, since high order contributions to the $N$ particles states can be important, but certainly $N \le M(n):=n(\tilde{N}-1)+1$, where $\tilde{N}$ is the number of particles created or annihilated at each interaction. In Yang-Mills field theory $\tilde{N}=2$.
\footnote{Here diagrams with four external lines are not considered, since they give a non renormalizable contribution. However, when all diagrams are summed up, the contribution of the contact interaction is null.} 

We can introduce states of well defined number of effective partons, $\ket{\psi_N}$, so that
\begin{equation}
 \ket{\Psi_n}=\sum_{N} \braket{\psi_N|\Psi_n} \ket{\psi_N}\,. 
\end{equation}
The states $\ket{\psi_N}$ are autovectors of $H_o$ with fixed number of particles, $N$. Of course $\braket{\Psi_n|\psi_N}=0$ whenever $N > M(n)$, and $\braket{\Psi_{N'}|\Psi_N}=\delta_{N'N}$.

Since the number of partons is fixed and they do not interact  but by contact interaction, the states $\ket{\psi_N}$ can be understood as the states of an ideal gas of $N$ partons. Therefore
\begin{equation}
 \ket{\psi_N}={\cal S} \ket{\gamma_1,m_1,p_1, \dots , \gamma_N,m_N,p_N}\,,
\end{equation}
where $m_i$ and $p_i$ are the mass and momentum of the $i$ partonic state, and $\gamma_i$ represents all relevant quantum numbers necessary to completely characterize the partonic state. ${\cal S}$ is the symmetrization operator acting over fermions and bosons. Since the mass of the effective partons varies continuously, ${\cal S}$ gives a negligible modification of the single parton states, so mass and momentum of each parton can vary independently, as far as the total energy is conserved.

Notice that the states with $N$ partons can be obtained in several ways, since
\begin{equation}
 \ket{\Psi(t)}=\sum_n  \sum_{N} \braket{\Psi_n|\Psi(t)} \braket{\psi_N|\Psi_n} \ket{\psi_N} \,,
\end{equation}
with $\braket{\psi_N|\Psi_n} \ne 0$ for $M(n)\geq N$.  For sufficiently high number of interactions, $n$, there are so many ways to obtain a $N$ particle system that the possibility to get a particular configuration becomes insensitive to the initial state. This situation is similar to that of complex systems where a statistical approach is possible. Therefore we assume that the probability to get a particular state can be calculated on the sole basis of the number of configurations corresponding to that state, and supposed that all configurations have the same probability to be observed. This assumption is not really new, since it is in fact the same statistical assumption made in LQCD where the gluon field configuration is obtained by statistical approach following a Monte Carlo method.

The probability to find a state where at least one parton has mass between $m_o$ and $m_o+dm_o$, and momentum coordinates  between $p_{oi}$ and $p_{oi}+dp_{oi}$ is given by matrix elements like
\begin{equation}
 \braket{\gamma_o,m_o,p_o, \dots|\Psi(t)}=\sum_n  \sum_{N} \braket{\Psi_n|\Psi(t)} \braket{\psi_N|\Psi_n} \braket{\gamma_o,m_o,p_o,\dots|\psi_N}\,, \label{partonfokstate}
\end{equation}
so we have the probability $P(m_o,p_o)d^4p_o=\braket{\gamma_o,m_o,p_o, \dots|\Psi(t)}$.

For reasons that will become apparent later, we will suppose that the state $\ket{\Psi(t)}$ has energy between $E$ and $E+dE$ that follows a probability distribution $P(E)$. The first bracket in Eq.~(\ref{partonfokstate}) is related to the probability to have $n$ interactions up to the instant $t$, and depends on the intensity of the interaction, $G$, so
\begin{equation}
 \braket{\Psi_n|\Psi}=G^n P(E) dE\,. \label{PnE}
\end{equation}
Therefore, the bracket in Eq.~(\ref{partonfokstate}) must be associated to the probability $P(m_o,p_o,E)d^4p_odE=\braket{\gamma_o,m_o,p_o, \dots|\Psi(t)}$.

The second bracket depends on the relative number of possibilities to get the configuration with $N$ particles after $n$ interactions, so
\begin{equation}
 \braket{\psi_N|\Psi_n} = C_N(n)\, \label{PNn}
\end{equation}
with
\begin{equation}
 \sum_{n} C_N(n)=1 \,. 
\end{equation}
The last bracket in the expression above is calculated statistically, see appendix for technical details. The result is
\begin{equation}
f(p_j) d^4p_j =  \frac{1}{8\pi}\frac{\Gamma(4N)}{\Gamma(4(N-1))} E^{-4} \left( 1 - \frac{p_j^0}{E} \right)^{4N-5} d^4p_j \,. \label{PNdistrib}
\end{equation}

Notice that the density of states with one particle with energy $p^0_j$ can be obtained by integrating on the vector momentum coordinates, resulting in
\begin{eqnarray}
f(p_j^0) dp_j^0 &=&   \frac{1}{6}\frac{\Gamma(4N)}{\Gamma(4(N-1))} E^{-1} \left( \frac{p_j^0}{E}\right)^3 \left( 1 - \frac{p_j^0}{E} \right)^{4N-5} dp_j^0  \,. \label{P0Ndistrib}
\end{eqnarray}
Observe that if we write $E=(4N-5)\mu$, and take the limits $N \rightarrow \infty$ and $\mu \rightarrow 0$ keeping constant $N\mu=kT$, we obtain the exponential factor usually found in the thermodynamical limit. Here, however, we cannot take such limit because $N<M(n)$.

At the rest mass frame the total energy of the system is equal to its mass, i.e., $E = M$,  and setting $p^0_j=\varepsilon_j$, then Eq.~(\ref{PNdistrib}) can be written as 
\begin{equation}
f(p_j) d^4p_j =  A(N) P_N\left( \frac{\varepsilon_j}{M} \right)  d^4\left( \frac{p_j}{M} \right) \,,
\end{equation}
with
\begin{equation}
 P_N\left( \frac{\varepsilon_j}{M} \right) = \left( 1 - \frac{\varepsilon_j}{M} \right)^{4N-5}\,,
\end{equation}
and
\begin{equation}
 A(N)=\frac{\Gamma(4N)}{8\pi\Gamma(4(N-1))}\,.
\end{equation}
From the expression for $A(N)$ we see that the number of states increases with $N^4$, so those configurations with large number of particles are favoured. The maximum number possible is $M(n)=n(\tilde{N}-1)+1$, so the probability to get a configuration with $N$ particles will be, for sufficiently large $n$, approximately $(N/(n\tilde{N}-n))^4$, so using Eqs.~(\ref{PnE}) and~(\ref{PNn}) we obtain

\begin{equation}
 P(m_o,p_o,E)d^4p_odE= \sum_n  \sum_{N} G^n
  \left(\frac{N}{n\tilde{N}-n}\right)^4 \left(1-\frac{\varepsilon }{M }\right)^{4N-5} d^4\left(\frac{p }{M }\right) P(E)dE\,. 
\end{equation}

Observe that when $N$ is sufficiently large and $\varepsilon/M$ sufficiently smaller than unit, we have
\begin{equation}
    \left[1-\frac{\varepsilon}{M}\right]^{(4N-5)}\left[1+\frac{\varepsilon}{M}\right]^{(4N-5)}=\left[1-\frac{ \varepsilon^2}{M^2}\right]^{(4N-5)}\sim 1\,.
\end{equation}
therefore we can set
\begin{equation}
     \left[1-\frac{\varepsilon}{M}\right]^{(4N-5)}=\left[1+\frac{\varepsilon}{M}\right]^{-(4N-5)}
\end{equation}
and finally obtain
\begin{equation}
 P(m_o,p_o,E)d^4p_odE= \sum_n  \sum_{N} G^n
  \left(\frac{N}{n\tilde{N}-n}\right)^4 \left(1+\frac{\varepsilon }{M }\right)^{-(4N-5)} d^4\left(\frac{p }{M }\right) P(E)dE\,. \label{probnonscale}
\end{equation}

\section{Self-similarity and fractal structure}

The result obtained in Eq.~(\ref{probnonscale}) shows that for an ideal gas with finite number of particles, the probability depends on a power-law function of the ratio $\varepsilon_j/M$. While this is valid for an ideal gas, where particles have no internal structure, the same result cannot be directly applied to the gas of effective partons because  effective partons always have internal structure related to self-energy contributions coming from interaction with the vacuum. We will now investigate how the scaling properties determined by the renormalization group equation can be used to obtain the result for effective partons.

%

The scaling properties represented by the Callan-Simanzyk Equation demands that effective partons at any scale are similar to any other parton after the proper rescaling is performed. In other words, if partonic properties are expressed in terms of scale free variables they must be described by the same functions of those variables. The energy distribution of a parton, as given by Eq.~(\ref{probnonscale}), depends on the ratio $\chi=\varepsilon_j/E$. Now let us consider that the system with energy $E$ in which the parton with energy $\varepsilon_j$ is one among $N$ constituents, is itself a parton inside a larger system with energy $\cal{M}$. Then the ration $E/{\cal M}$ is represented by the same variable, $\chi$, that describes the ration $\varepsilon_j/E$. In addition, we can write $P(E)$ in Eq.~(\ref{probnonscale}) as $P(E/{\cal M})$, and this probability density must follow that same function as $P(\varepsilon_j/E)$, that is
\begin{equation}
 P\left(\frac{\varepsilon_j}{E}\right)\sim P\left(\frac{E}{{\cal M}}\right) \sim P\left(\chi \right)\,, 
\end{equation}
where the sign $\sim$ indicates that the three functions above are indeed the same function of the scale free variable $\chi$. The same conclusion is valid for any parton with energy $\varepsilon$ which is a component of another parton with energy $\Lambda$, and to let it clear in the following calculations we write $\chi=\varepsilon/\Lambda$. The reasoning just used here is the same that applies to fractal structures, so what we are doing is to introduce the mathematical tools common to fractals studies in the analysis of Yang-Mills fields.

The self-similarity among the partons implies that the probability that the parent parton with mass $E$ inside a larger system with mass ${\cal M} $ is similar to the probability given in Eq.~(\ref{probnonscale}), so the dependence on $N$ that appears in the exponent must be changed to a parameter $\alpha$, which remains to be determined. This parameter represents to the total number of degrees of freedom of the fractal, playing the same role of the exponent $4N-5$ in the case of the ideal gas.  We can write
\begin{equation}
 P\left(\frac{E}{{\cal M}}\right)=\left(1+\frac{E}{{\cal M}}\right)^{-\alpha} \,,
\end{equation}
and the function being integrated in Eq.~(\ref{probnonscale}) contains the term
\begin{equation}
  \left[P\left(\frac{E}{{\cal M}}\right)\right]^{\nu} P_N\left(\frac{\varepsilon }{E}\right)  d\left(\frac{\varepsilon }{E}\right) \sim \left[1+\frac{\varepsilon}{\Lambda}\right]^{-\alpha \nu} \left[1+\frac{\varepsilon}{\Lambda}\right]^{-(4N-5)} d\left(\frac{\varepsilon}{\Lambda}\right)\,. \label{recurrenceformula}
\end{equation}
Notice that the parameters $\alpha$ and $\nu$ describe the complexity of the interaction involved in the gauge field interaction, and in this sense they measure the sensibility of the effective parton to its internal degrees of freedom.

The self-similarity implies, from relation~(\ref{recurrenceformula}), that
\begin{equation}
 (4N-5)+\alpha \nu=\alpha\,,
\end{equation}
since the same probability governs any parton distribution.Therefore
\begin{equation}
 \alpha=\frac{4N-5}{1-\nu}\,.
\end{equation}
The parameter $\nu$ represents the fraction of total number of degrees of freedom of the state $\ket{\psi_N}$ that is involved in each interaction.
Observe that we expect $\nu \leq 1$, therefore $\alpha$ is positive.

Observe that $q$ is related to the resolution parameter $\nu$ and to the number of particles, $N$. But this two parameters are not independent, since $N$ increases as the resolution increases. In fact, we can write
\begin{equation}
 \Lambda=\alpha \lambda\,,
\label{resolution}
\end{equation}
where $\lambda$ is a reduced scale, independent of the number of degrees of freedom relevant to the system. Since variable $\chi$ must follow a universal distribution for any parton that is independent of the level in the fractal structure, and since the smallest parton is the one obtained in  1-loop approximation, where $N=\tilde{N}$, and therefore is constant, then also $1/\alpha=\lambda/\Lambda$ is independent of the position the system occupy in the fractal structure, and so $\alpha$ is constant.  We can introduce a parameter $q$ such that
\begin{equation}
 \frac{1}{\alpha}=q-1\,,
\label{qvalue}
\end{equation}
with $q>1$ been constant for any parton in the fractal system.
Then we obtain
\begin{equation}
 P(\varepsilon/\lambda)=\left[1+(q-1)\frac{\varepsilon}{\lambda}\right]^{\frac{-1}{q-1}}\,. \label{Tdistribution}
\end{equation}

This result shows that the distribution of parton energy created by a system governed by Yang-Mills fields depends only on the ratio between the parton energy, $\varepsilon$, and the energy scale per degree of freedom $\lambda$. Furthermore, it shows that the energy distribution follows the q-exponential function commonly found in Tsallis non extensive statistics. 

This result shows that the distribution of parton energy created by a system governed by Yang-Mills fields depends only on the ratio between the parton energy, $\varepsilon$, and the energy scale per degree of freedom $\lambda$. Furthermore, it shows that the energy distribution follows the q-exponential function commonly found in Tsallis non extensive statistics. 

Similar results have been obtained through a different approach using the concept of thermofractals, introduced in Ref.~\cite{Deppman2016} and studied in details in Ref.~\cite{DFMM}. There it is shown that the fractal structure leads to the non extensive statistics, and it is discussed the relations between thermofractals and Hagedorn's self-consistent thermodynamics developed to study high energy collisions~\cite{Hagedorn0, Hagedorn65}, and that was extended to non extensive statistics~\cite{Deppman2012}.

The results obtained here mean that an effective parton with energy $\Lambda$ can be represented as a system of effective partons at an energy scale  $\lambda$, which can be represented by states of the type $\ket{\gamma_1,\varepsilon_1, \dots,\gamma_N,\varepsilon_N}$ with $\varepsilon_1, \dots, \varepsilon_N \sim \lambda$. A particular state where one particle has energy between $\varepsilon_o$ and $\varepsilon_o+d\varepsilon_o$, denoted by $\ket{\gamma_o,\varepsilon_o}=\ket{\gamma_o,\varepsilon_o,\dots}$, and the matrix element becomes
\begin{equation}
 \braket{m_o|\Psi(t)}=\tilde{N}^r \left[1+(q-1)\frac{\varepsilon_o}{\lambda}\right]^{-1/(q-1)}\,.
\end{equation}
This is obtained by increasing one order in the perturbative calculation, that is, this results from a vertex function $Z$ to which one loop is added and an external line is added to the new loop. This means that the q-exponential plays the role of an effective coupling constant in the vertex function,
that is
\begin{equation}
 Z=\textrm{Tr}\braket{\Psi_{n+1}| g e^{iH_ot_{n+1}}|\Psi_n}
\end{equation}
with
\begin{equation}
 g=\prod_{i=1}^{\tilde{N}}G\left[1+(q-1)\frac{\varepsilon_o}{\lambda}\right]^{-1/(q-1)}\,. \label{effectivecoupling}
\end{equation}

\begin{figure}[t!]
 \centering
   \begin{subfigure}[]
        \centering
        \includegraphics[scale=0.5]{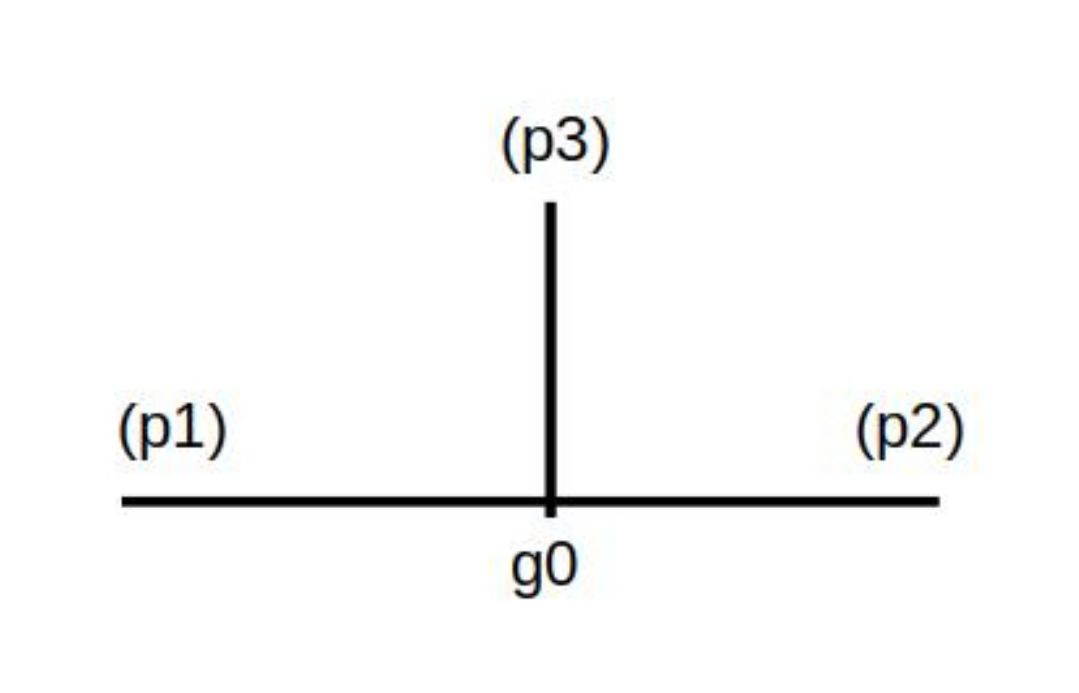}
   \end{subfigure}%
    ~ 
   \begin{subfigure}[]
        \centering
        \includegraphics[scale=0.5]{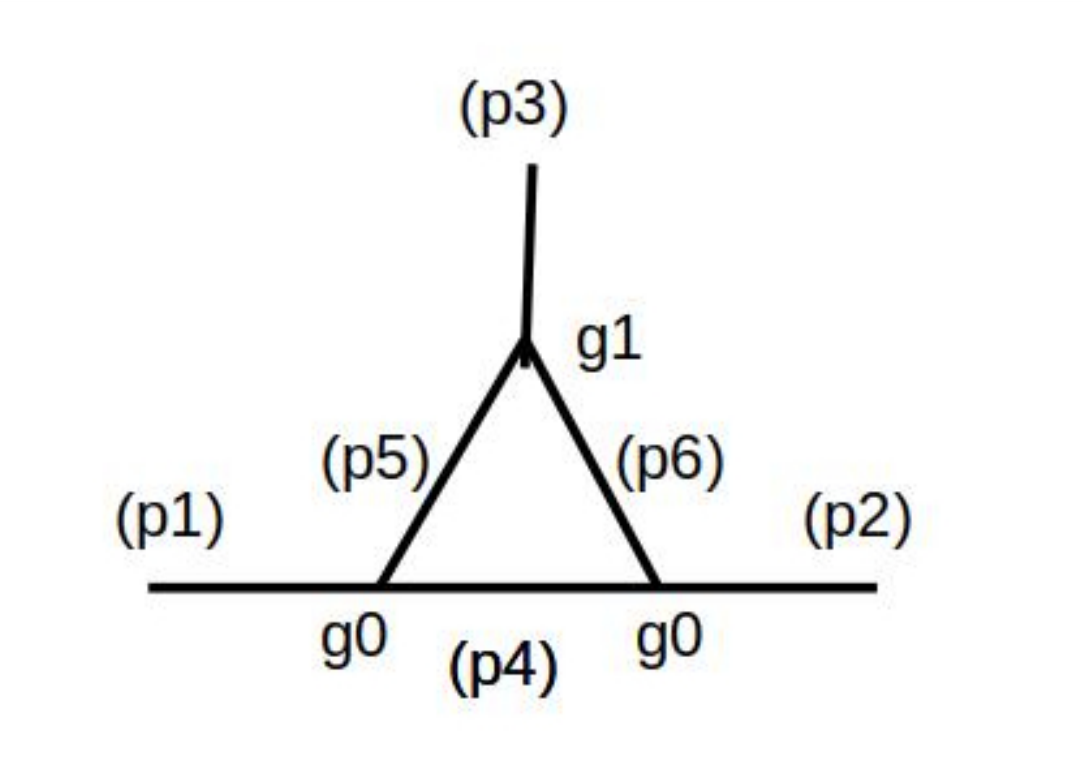}
        \caption{Vertex functions at scale (a) $\lambda_o$ and (b)$\lambda$.}\label{scalignvertexes}
   \end{subfigure} 
\end{figure}

\section{Discussion and applications}

As an application of the formalism proposed here, let us consider a vertex in two different orders, as depicted in Fig.~\ref{scalignvertexes}. The vertex function is
\begin{equation}
 \Gamma_o=\braket{\gamma_2 p_2 \gamma_3 p_3|g(\lambda_o)e^{iH_ot}| \gamma_1 p_1}\,.
\end{equation}
 The next order in perturbative approximation is given by the vertex with one additional loop, which results in a vertex function
 \begin{equation}
  \Gamma=\braket{\gamma_2 p_2 \gamma_3 p_3|\bar{g}e^{iH_ot}| \gamma_1 p_1}\,,
 \end{equation}
where $\bar{g}$ is the effective coupling given by
\begin{equation}
 \bar{g}=g(\lambda_o) e^{iH_ot_3}\ket{\gamma_2 p_6,\gamma_3 p_3, \gamma_4 p_4} \Gamma_M \bra{\gamma_1 p_5, \gamma_4 p_4}g(\lambda_o)\,,
\end{equation}
with
\begin{equation}
 \Gamma_M=\braket{\gamma_2 p_6, \gamma_3 p_3, \gamma_4 p_4|g(\lambda)e^{iH_ot_2} | \gamma_1 p_5, \gamma_4 p_4}\,.
\end{equation}

Observe that the effect of the effective coupling as defined by Eq.~(\ref{effectivecoupling}) is to increase the contribution of low energy partons and decrease the contribution of high energy partons. At each vertex, the creation of a parton with energy much higher than that expected for partons at the fractal level as determined by the scale $\lambda$ is strongly suppressed. The same is valid for the parton annihilation: partons with high energy are more likely to remain in the system with small probability to interact with the less energetic and therefore less massive partons in the medium. Depending on its color content, the heavy parton can reach the surface of the medium and scape. In this way, a complete description of the behaviour of those heavy partons must include the possibility of color exchange and coalescence, and some works in this direction, using power-law distributions, have already been done~\cite{Feal}. Keeping these aspects in mind, we continue here with a purely cinematic description of the partonic evolution of QCD. Due to energy-momentum conservation constraint, $\varepsilon_5+\varepsilon_6\sim \lambda$, so the coupling strongly favors configurations where $\varepsilon_5 \sim \varepsilon_6 \sim \lambda/2$. In this way the effective coupling controls how the energy transferred to an effective parton is distributed among its internal degrees of freedom.

Notice that $\Gamma_M$ and $\Gamma_o$ are similar, differing only by the presence of the non interacting field $(\gamma_4 p_4)$ and by the different energies of the fields at each vertex. The scaling properties of Yang-Mills fields allow us to relate $\Gamma_M$ to $\Gamma_o$ by an appropriate scale, $\lambda$. This scale must take into account that this is a 1-loop contribution, so $\lambda/\lambda_o=\tilde{N}^{-1}$, and the coupling constant and fields in $\Gamma_M$ must be modified accordingly to~\cite{Dyson1,Gell-Mann_Low}
\begin{equation}
 \phi_i(p,\bar{m})=Z_i^{-1}\phi_i(p,m)\,,
\end{equation}
where $m$ is the mass of the parton at scale $\lambda_o$ and $\bar{m}$ is its mass at scale $\lambda$. Here $\phi_i$ refers to the field of parton $i$, and $Z_i$ are multiplicative factors arising from the vertex renormalization.

The renormalization properties imposes that~\cite{Dyson1, Gell-Mann_Low}
\begin{equation}
 Z_i^{-1}=1+\int_{\lambda_o}^{\lambda} \gamma_i(m/\lambda, \bar{g}) \frac{d\lambda}{\lambda}\,.
\end{equation}
For one loop approximation we have $\lambda=\lambda_o+d\lambda$, where $d\lambda<0$, and the integration above turns into
\begin{equation}
 Z_i^{-1}=1+\frac{g^2}{16\pi^2}\gamma_i\log(\lambda/\lambda_o)\,, \label{Zi}
\end{equation}
from where it follows that
\begin{equation}
 \lambda \frac{\partial \phi_i}{\partial \lambda}=\gamma_i\phi_i\,. \label{gammai}
\end{equation}
The field $(\gamma_4 p_4)$ in $\Gamma_M$ does not interact at this vertex, so it does not need to be scaled, so it will be omitted in the following calculation for the sake of clarity.

The scaling behavior of the $\Gamma(M)$ function is obtained from dimensional analysis, and it results to be
$\Gamma_M(\lambda)=(\lambda / \lambda_o)^4$, after energy-momentum conservation is taken into account,
so
\begin{equation}
 M \frac{\partial \Gamma}{\partial \lambda}=d \,\,\Gamma\,,
\end{equation}
with $d=4$ in the present case.
From these considerations and from Callan-Symanzik equation, it results that
\begin{equation}
 \beta_{\bar{g}}\frac{\partial \Gamma}{\partial g}=-(d+\gamma_5+\gamma_6)\Gamma\,.
\end{equation}

The contribution to the scaling transformation from $\bar{g}$ appears through $g(\lambda)$, while the other two vertex are at the inital scale, that is, $g=g(\lambda_o)$. Also, in order to compare with QCD results, we study the behavior of $g(\lambda)$ at $\lambda=\lambda_o/\mu$. We have the logarithmic derivative
\begin{equation}
  \mu \frac{\partial g(\mu)}{\partial \mu}= -\sum_{j=5,6} g  \mu\,\varepsilon_j \left[ 1+(q-1) \mu \varepsilon_j/\lambda_o \right]^{-1}\,. \label{betafunc}
\end{equation}
The beta function can be easily calculated 
and we get
\begin{equation}
 \beta_{\bar{g}}=-\frac{1}{16\pi^2} \frac{1}{q-1} g^{\tilde{N}+1}\,,
\end{equation}
and we emphasize that $\tilde{N}=2$.

Scaling properties of QCD have been extensively studied in the 1-loop approximation. The beta-function for QCD is~\cite{Gross:1973id,Politzer:1974fr}
\begin{equation}
 \beta_{\QCD}=- \frac{g^3}{16\pi^2}  \left[\frac{11}{3}c_1-\frac{4}{3}c_2\right]\, 
\end{equation}
where
\begin{equation}
  c_1 \delta_{ab}=f_{acd}f_{bcd}
\end{equation}
and
\begin{equation}
  c_2 \delta_ab=\textrm{Tr} (T_a T_b)\,,
\end{equation}
therefore relating the entropic index, $q$, from Tsallis statistics to fundamental parameters of the field theory. 

Quantitatively, the parameters $c_1$ and $c_2$ are related to the number of colors and flavors by $c_1=N_c$ and $c_2=N_f/2$. Using $N_c=N_f/2=3$ we get
\begin{equation}
  \frac{11}{3}c_1-\frac{4}{3}c_2=7\,,
\end{equation}
which leads to $q=1.14$. From experimental data analysis\cite{Lucas, Lucas2, Sena} we have $q=1.14 \pm 0.01$, showing a good agreement between theory and experiments. 

\section{The beta function and the effective coupling}

\begin{figure}[t!]
 \centering
   \begin{subfigure}[]
        \centering
        \includegraphics[scale=0.45]{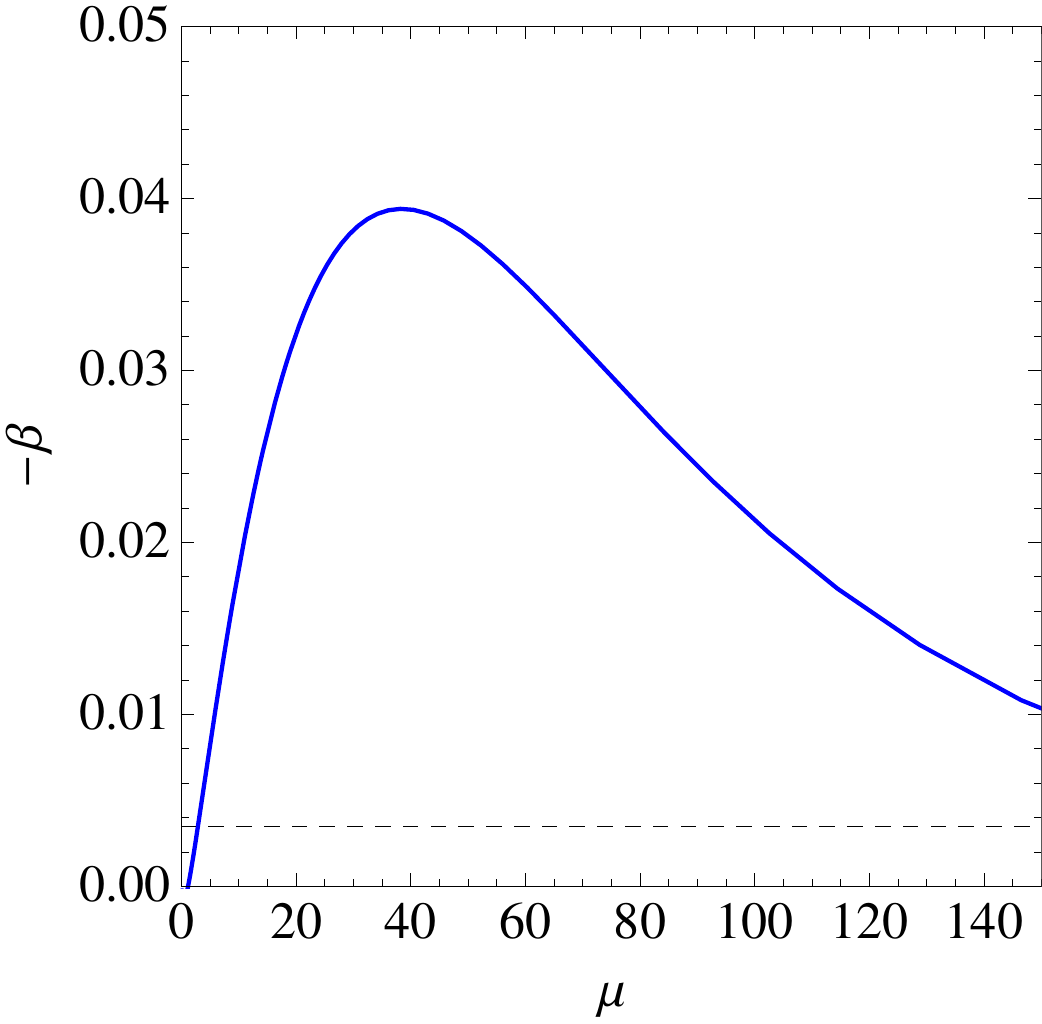}
   \end{subfigure}%
    ~ 
   \begin{subfigure}[]
        \centering
        \includegraphics[scale=0.45]{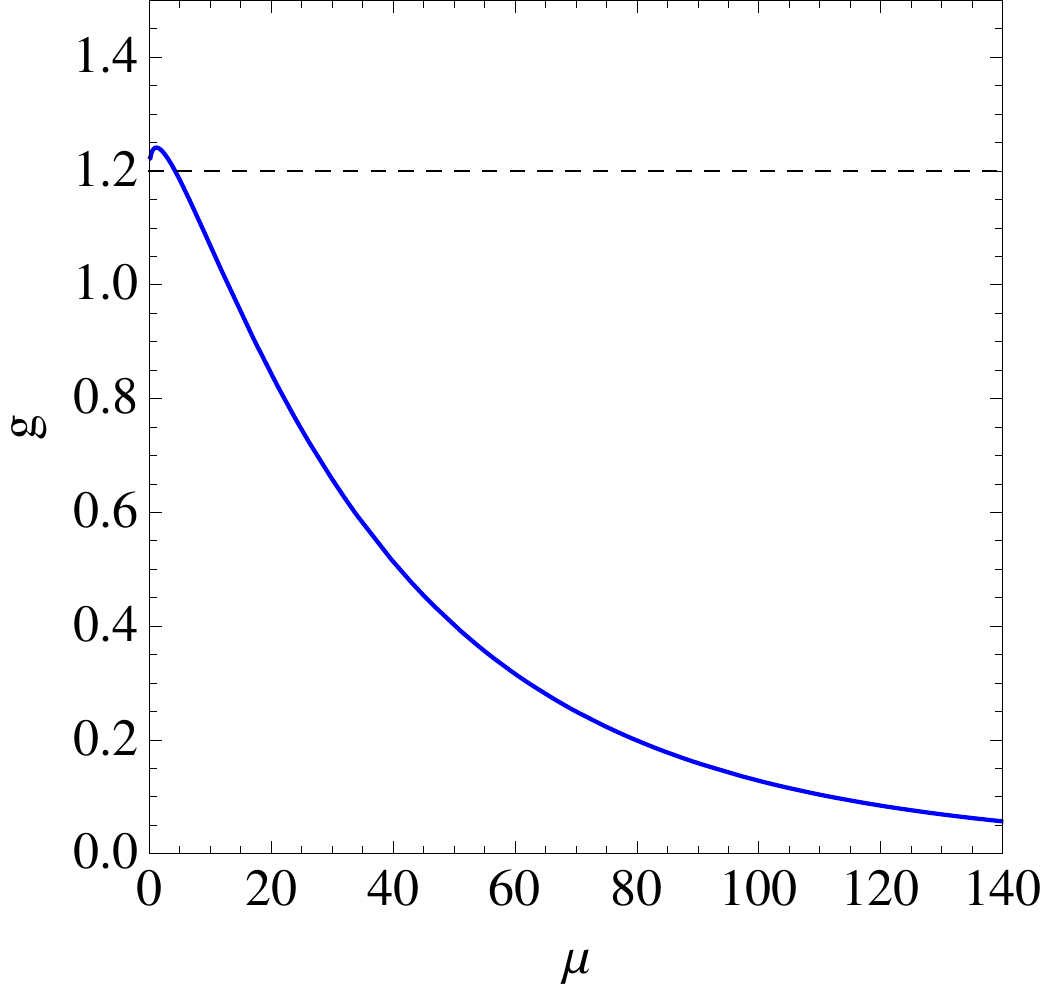}
   \end{subfigure} 
   \begin{subfigure}[]
        \centering
        \includegraphics[scale=0.45]{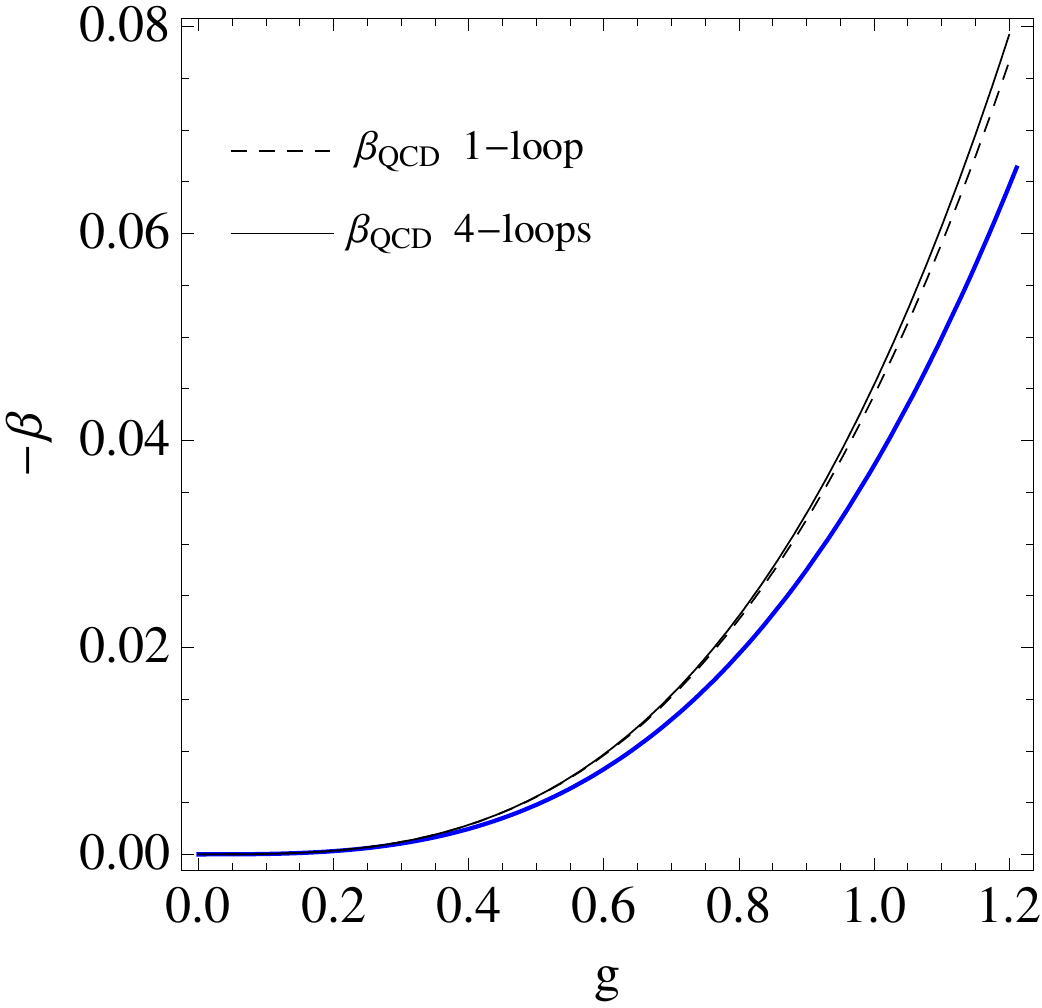}
   \end{subfigure}    
        \caption{Behavior  of (a) the logarithmic derivative of the effective coupling with respect to the scale $\mu$, (b) effective coupling, and (c) beta function against effective  coupling. The values $G=$3.67 and $\varepsilon_5=\varepsilon_6=\lambda/2$ were used in the computation of plots (a) and (b), and $G=752$ in (c). The dashed line represents the  value from QCD in one-loop approximation~\cite{Gross:1973id,Politzer:1974fr}, and the dash-dotted line represents $\beta$ as calculated in 4-loop approximation~\cite{vanRitbergen:1997va,Czakon:2004bu}.}\label{gg'}
\end{figure}

The result obtained here shows that in the 1-loop approximation we recover the behavior expected for QCD. Let us now consider the asymptotic limit, when $\mu \rightarrow \infty$. This limit is equivalent to say that the 1-loop approximation is made at higher orders of logarithmic approximation, i.e., when $\lambda_o\rightarrow \lambda'_o \ll \lambda_o$. In this case also $\lambda \rightarrow \lambda'$, such that the ratio $\lambda'/\lambda'_o \sim \lambda/\lambda_o$, i.e., we are doing the same 1-loop approximation, but to a process that occurs at higher order of logarithmic approximation. At this point, we have to take into account that the fractal structure considered here, with all effective parton showing an internal structure, is broken since quarks are elementary particles. This implies  the existence of an inferior limit to $\varepsilon=\varepsilon_o$. When doing so, we have a minimum value for the scale, $\lambda_c=(q-1)\varepsilon_o$, then the scale is redefined to $\lambda'/\mu=\lambda_o/\mu-\lambda_c$. 
It results that $g(\mu') \rightarrow 0$ and also the logarithmic derivative
, $\mu\partial g(\mu)/\partial \mu \rightarrow 0$, as shown in Fig.~\ref{gg'}. These results confirm that the effective coupling used here leads to the asymptotic freedom of quarks obtained in QCD, since $g \rightarrow 0$ as $\mu \rightarrow \infty$. The beta function also approaches zero asymptoticaly, but for low values of the scale parameter $\mu$ it also goes to zero, presenting a peak around $\mu/\lambda_o=40$. For a direct comparison with the QCD running coupling, we show in Fig.~\ref{gg'} the comparison with QCD calculations in 1-loop and in 4-loop calculations. We observe that in the asymptotic region, when $g \rightarrow 0$, the behavior of the effective coupling proposed in the present work is similar to the QCD behavior, diverging from it for larger coupling values, when perturbative QCD is not expected to work.

 It is interesting to note that the scaling invariance can be used in two ways: one can fix a value for $\lambda$, based on experimental resolution or some physical change in the system that makes the scale invariance invalid or useless, and then the several fractal layers are obtained by considering systems of large and large energies corresponding to different values of $\Lambda$; the other way is to fix the total energy and let $\lambda$ vary, meaning that the system is observed with better resolution, and therefore deeper layers of the system are investigated, i.e., the better the resolution, the lower the $\lambda$. The first way finds application in high energy collisions, and the second, in studies of hadron structure.

The results obtained here have shown that a system with fractal structure, similar to the thermofractals introduced in Ref~\cite{Deppman2016} and studied in~\cite{DFMM}, can be understood as a natural consequence of the scale invariance of gauge field theories. There is is shown that the fractal structure leads to the non extensive statistics, and it is discussed the relations between thermofractals and Hagedorn's self-consistent thermodynamics developed to study high energy collisions~\cite{Hagedorn0, Hagedorn65}, and that was extended to non extensive statistics~\cite{Deppman2012}.
This fractal structure has been already used to investigate properties of hadrons~\cite{PedroCardoso}, phase-transition in hot hadronic matter~\cite{Megias} and  neutron stars~\cite{Debora}. The power-law distribution of energy and momentum, which is a direct consequence of the fractal structure, was used to describe $p_T$ distributions from high energy collisions experiments~\cite{Lucas, Lucas2, WilkWlodarczyk} and to describe hadron mass spectrum~\cite{Lucas,Deppman_Universe}. The results obtained here give a stronger basis for the interpretation of those experimental and phenomenological studies in terms of non extensive statistics.

\section{Multiplicity}
It is possible to understand, from the considerations made here, that the fractal structure of YMF is the basis for investigations of hadron properties~\cite{PedroCardoso}, phase-transition in hot hadronic matter~\cite{Megias}, neutron stars~\cite{Debora} and cosmic-ray~\cite{BeckCosmicRay}.  These phenomenological approaches are, in fact, implementations of  scaling symmetries observed in Yang-Mills fields. The fractal structure also allows the understanding of the self-similarity~\cite{WWselfsymmetry, Tokarev, Zborovsky} and scaling properties observed in high energy experimental data. In fact, these findings are direct consequences of the scaling properties of YMF, as discussed here.
Moreover, the fact that the entropic index, $q$, is obtained from well-known field-theoretical parameters, the results we have obtained allows a new interpretation of Tsallis statistics in terms of fractal structure in the same lines it was obtained in thermofractals approach~\cite{Deppman_Universe}.

The fractal structure presents at least one fractal dimension, and the Haussdorf dimension is a characteristic dimension that can be calculated by using the box-counting technique~\cite{Falconer}, where the dimension $D$ is related to the number of boxes, ${\cal N}$, necessary to completely cover all possible values for the measured quantity and $D_t$ is the topological dimension. At some scale $r$ these quantities are related by\cite{Falconer}
\begin{equation}
    {\cal N} r^{-D} \propto r^{-D_t}\,.
\end{equation}
In our case, $D_t=1$ since we are dealing with system energy as a measure. The procedure to obtain the Haussdorf dimension is similar to that followed in Ref.~\cite{Deppman2016}. The {\bf average} energy of the partons at  the scale $\lambda$,  already  introduced as the energy scale per degree of freedom, is
\begin{equation}
 \braket{\varepsilon}= \int_0^{\infty} \varepsilon P(\varepsilon) d\varepsilon=\frac{\lambda}{2q-1}\,.\label{eav}
\end{equation}
The ratio between the average energy of the components and the parent system energy
\begin{equation}
 R=\frac{\braket{\varepsilon}}{E} \label{EqR}
\end{equation}
is related to the level of the fractal structure relative to the scale $\lambda$ by
\begin{equation}
 R^n=\frac{\lambda}{E}=r\,. \label{Rr}
\end{equation}
The number of boxes with length $\lambda$ necessary to completely cover the possible range of energies in which the fractal components can be found is ${\cal N}=\tilde{N}^n$.
Then it follows from Eq.~(\ref{Rr}) that
\begin{equation}
 n=\frac{\log r}{\log R}\,. \label{Eqn}
\end{equation}

Eq.~(\ref{Rr}) also shows that, in terms of the scales, the energy of the system varies as $E\sim r^{-1}$. Let us now write the dependence of the parton energies at scale $\lambda$ as $\varepsilon \sim r^{-D}$, then
\begin{equation}
 {\cal N} r^{-D} \propto r^{-1}\,,
\end{equation}
therefore
\begin{equation}
 D-1=n \frac{\log \tilde{N}}{\log r}\,.
\end{equation}
From Eq.- (\ref{Eqn}) it follows that
\begin{equation}
 D-1=\frac{\log\tilde{N}}{\log R}\,. \label{EqD}
\end{equation}

From Eqs.~(\ref{eav}) e~(\ref{EqR}), and using $E=\lambda_r/(q-1)$, we get
\begin{equation}
 R=\frac{q-1}{2q-1}\,.
\end{equation}
Using the value $q-1=1.14$ it follows that $D=0.69$. This result is in good agreement with the fractal dimension observed in intermittency analyses of high energy experimental data~\cite{Bialas_Peschanski, Bialas_Peschanski2}. These analyses allows to access fractal dimensions by studying the behavior of cummulants of the measured distributions~\cite{Hwa, HwaPan, Hegyi1, DreminHwa, Hegyi2, Antoniou, Wolf, Kittel }, and the systematics show that for $pp$ collisions there is a good agreement between the value obtained from the theory with those resulting from experimental data analyses.

The fractal dimension gives the behavior of the parton energy with the energy scale, $r$, that is, while the total energy goes as $E \propto r^{-1}$, the partons observed at scale $\lambda$ have energies that depend on the scale as $\varepsilon \propto r^{-D}$. A more direct way to access the fractal dimension is the particle multiplicity. In fact, being $\cal{M}$ the particle multiplicity, we have 
\begin{equation} 
  {\cal M} \braket{\varepsilon}=E \,. 
\end{equation} 
From the dimensional behavior obtained above, we get 
\begin{equation} 
  {\cal M}=E^{1-D}\,. 
\end{equation} 
For the case of hadrons, as we have seen, $q=1.14$ and $D=0.69$, so we obtain ${\cal M}\propto E^{0.31}$, which is in excellent agreement with the result obtained for $pp$ collision at high energy~\cite{Sarkisyan_Multiplicity}, which gives, for a power-law fit, an exponent corresponding to $1-D=0.302$. 
\section{Conclusions}

In the present work the scaling properties f Yang-Mills fields are analyzed under the light of the concept of fractals. It is shown that those scaling properties lead to the formation of a fractal structure, and usig a thermodyanamical hypotheis similar to that used in LQCD we obtain Tsallis distributions that have been associated to the long-tail distributions observed in high energy data. The entropic index, $q$, which in Tsallis statistics is in most cases a parameter to be determined experimentally, in the present case can bedetermined completely in terms of the fundamental parameters of the field theory.

The result is used to obtain, by a recurrence formula that reflects the self-similar features of the fractal, the effective coupling which is expressed in terms of a scale dependent formula. Applying this expression for the case of QCD in the 1-loop approximation, the entropic index is calculated, for the first time, from the numbers of colors and flavors. The result is shown to be in good agreement with the value obtained for $q$ by fitting Tsallis distributions to data.
From the analysis of the fractal structure we obtain the fractal dimension associated to Yang-Mills fields, which is determined completely in terms of the field theory parameters. The fractal dimension is calculated for QCD and the result is in good agreement with the value obtained from intermittency analysis of high energy distributions.

Finally, it is shown that the fractal dimension allows us to determine the behaviour of the particle multiplicity against the collision energy, and we obtain a result that is in good agreement with that observed in high energy collisions.

In summary, we have shown in the present work that renormalizable field theories lead to fractal structures, which can be studied, from a thermodynamical point of view, with Tsallis statistics. A recursive method allows to perform non perturbative calculations to describe the particles structure governed by the gauge theory. In the case of multiparticle production, the calculations lead to a thermodynamical description where non extensive statistics must be used. The results obtained here give a solid basis from QCD to the use of non extensive self-consistent thermodynamics to describe properties of strong interacting systems and to the use of thermofractal structure to describe hadrons.

\bigskip

{\bf Acknowledgments:} 
A.D. would like to thank the University of Granada, where part of this
work has been done, for the hospitality and financial support under a
grant of the Visiting Scholars Program of the Plan Propio de
Investigaci\'on of the University of Granada. He also acknowledges the
hospitality at Carmen de la Victoria. A.D. and D.P.M.  are
partially supported by the Conselho Nacional de Desenvolvimento
Científico e Tecnológico (CNPq-Brazil) and by Project INCT-FNA
Proc. No. 464898/2014-5. The work of E.M. is supported by the Spanish MINEICO
and European FEDER funds under Grants FIS2014-59386-P and
FIS2017-85053-C2-1-P, by the Junta de Andaluc\'{\i}a under Grant
FQM-225, and by the Consejer\'{\i}a de Conocimiento, Investigaci\'on y
Universidad of the Junta de Andaluc\'{\i}a and European Regional
Development Fund (ERDF) Grant SOMM17/6105/UGR. The research of E.M. is
also supported by the Ram\'on y Cajal Program of the Spanish MINEICO
under Grant RYC-2016-20678.

\section{Appendix: Calculation of density of states}

Given a system with $N$ free particles with Hamiltonian
\begin{equation}
H = \sum_{i=1}^N p_i^0 \,,
\end{equation}
where $p_i^\mu = (p_i^0, \vec{p}_i)$ is the fourth momentum of particle $i$, the goal is to compute the density of states~$\rho(p)$, with the normalization condition
\begin{equation}
1 = \int d^{4N} p \, \rho(p) \,. \label{eq:norm}
\end{equation}

In the following we won't assume a fixed value for the mass $m_i$ of particle $i$,  where $m_i^2 = p^\mu p_\mu$, so that $p_i^0$ and $\vec{p}_i$ are variables that may change independently each other. Let us compute the phase space volume of region $H \le E$, i.e.
\begin{equation}
\Omega_N(E) = \int d^{4N}p \, \Theta\left( E- \sum_{i=1}^N p_i^0\right) \,,  \label{eq:Omega1}
\end{equation}
where $\Theta(x)$ is the step function. We can express the integral in the form
\begin{equation}
\Omega_N(E) = \int d^N p^0  \, \Theta\left( E- \sum_{i=1}^N p_i^0\right)  \prod_{i=1}^N \int d^3p_i \, \Theta(p_i^0 - |\vec{p}_i|) \,. \label{eq:Omega2}
\end{equation}

Note that for particle $i$ one has $\vec{p_i}^2 = (p_i^0)^2 - m_i^2$. Because $m_i$ is not fixed to a particular value,  the limit of the integration in the $d^3 p_i$ integral is $0 \le |\vec{p_i}| \le p_i^0$, and this has been expressed with the corresponding step function in Eq.~(\ref{eq:Omega2}). This integral can be easily computed to give $\int d^3p_i \, \Theta(p_i^0 - |\vec{p}_i|) = \frac{4}{3} \pi (p_i^0)^3$, so that Eq.~(\ref{eq:Omega2}) turns out to be
\begin{equation}
\Omega_N(E) = \left( \frac{4\pi}{3} \right)^N \int d^N p^0 \, \left( \prod_{i=1}^N (p_i^0)^3 \right)  \, \Theta\left( E- \sum_{i=1}^N p_i^0\right) \,.
\end{equation}

From dimensional analysis one can see that the result of this integral should be of the form
\begin{equation}
\Omega_N(E) = c_N \pi^N E^{4N} \,, \label{eq:OmegaNg}
\end{equation}
where $c_N$ are some coefficients to be determined, and the factor $\pi^N$ has been explicitly extracted for convenience. It is possible to obtain the coefficients $c_N$ by mathematical induction. In the case $N=1$, one can easily check that 
\begin{equation}
\Omega_1(E) = \frac{1}{3} \pi E^4 \,,
\end{equation}
so that $c_1 = 1/3$. Assuming that the expression of $\Omega_N(E)$ is known, let us compute $\Omega_{N+1}(E)$. It writes
\begin{eqnarray}
\Omega_{N+1}(E) &=& \int d^4p_{N+1} \int d^{4N}p \, \Theta\left(E - \sum_{i=1}^{N+1} p_i^0\right) = \int d^4p_{N+1}  \int d^{4N}p \, \Theta\left(E - p_{N+1}^0 - \sum_{i=1}^{N} p_i^0\right) \nonumber \\
&=& \int d^4 p_{N+1} \, \Omega_N(E - p_{N+1}^0)  \,,
\end{eqnarray}
where in the last equality we have performed the integral only in the momentum of particles $i=1, \cdots, N$. The integral in $d^4p_{N+1}$ can be easily performed following similar steps as above, i.e.
\begin{eqnarray}
\Omega_{N+1}(E) &=& \int d^4 p_{N+1} \, \Omega_N(E - p_{N+1}^0) = \int_0^E dp_{N+1}^0  \Omega_N(E - p_{N+1}^0) \int d^3p_{N+1} \, \Theta\left( p_{N+1}^0 - |\vec{p}_{N+1}| \right) \nonumber \\
&=& \frac{4\pi}{3} \int_0^E dp^0_{N+1} (p_{N+1}^0)^3 \, \Omega_N(E - p_{N+1}^0) \,.
\end{eqnarray}

Using now Eq.~(\ref{eq:OmegaNg}), this integral can be easily performed, and the result is
\begin{equation}
\Omega_{N+1}(E) = 8 \pi^{N+1}\frac{\Gamma(4N+1)}{\Gamma(4N+5)} c_N E^{4(N+1)}  \,.
\end{equation}
This result should be identified with $\Omega_{N+1}(E) = c_{N+1} \pi^{N+1} E^{4(N+1)}$, and from it one obtains a relation between $c_{N+1}$ and $c_N$, which can be used iteratively to obtain 
\begin{equation}
c_N = \frac{4! \cdot 8^{N-1}}{(4N)!} c_1 = \frac{8^N}{(4N)!} \,.
\end{equation}
In the last equality we have used that $c_1 = 1/3$. Finally, the result of Eq.~(\ref{eq:Omega1}) is
\begin{equation}
\Omega_{N}(E) = \frac{(8\pi)^N}{(4N)!} E^{4N} \,. \label{eq:ON}
\end{equation}

Let us introduce a constant distribution in the microcanonical ensemble
\begin{equation}
 \rho(p) =
\begin{cases}
 & C  \,, \qquad \textrm{for} \qquad E \le H \le E + \Delta E \\
 & 0 \,\,\,\,\,\,\,\,\,\,\,\,\,\,\,\,\textrm{otherwise}\,,
\end{cases}
\end{equation}
that should be normalized to $1$ as indicated in Eq.~(\ref{eq:norm}). Then, one finds that
\begin{equation}
1 = C \cdot \left[  \Omega_{N}(E + \Delta E) - \Omega_N(E) \right] \,,
\end{equation}
so that
\begin{equation}
C = \frac{1}{\Omega_{N}(E + \Delta E) - \Omega_N(E)} \,.
\end{equation}

The probability distribution for particle~$j$ to have momentum $p_j^\mu$ is obtained by integrating the joint distrubution $\rho(p_1, \cdots, p_N)$ over all the variables except $p_j$. Then, one has
\begin{eqnarray}
f(p_j) d^4 p_j &=& d^4p_j \int d^{4(N-1)} p \, \rho(p) \, \Theta( E \le  H \le E+ \Delta E ) \nonumber \\
&=& d^4 p_j \int d^{4(N-1)}p \, \rho(p) \, \Theta\left(E-p_j^0 \le \sum_{i=1}^{N'} p_i^0 \le E + \Delta E - p_j^0 \right) \,, 
\end{eqnarray}
where the prime in $\sum_{i=1}^{N'}$ means that the term $i=j$ should be excluded in the summation. Using the result above for $\rho(p)$, this can be expressed in the form
\begin{equation}
f(p_j) d^4 p_j = d^4 p_j \frac{\Omega_{N-1}(E + \Delta E - p_j^0) - \Omega_{N-1}(E - p_j^0)}{\Omega_{N}(E + \Delta E) - \Omega_N(E)} \,.
\end{equation}
Finally, by using the explicit expression of $\Omega_N(E)$ given by Eq.~(\ref{eq:ON}) and taking the limit $\Delta E \to 0$, one arrives at the final result
\begin{equation}
f(p_j) d^4p_j = d^4p_j \frac{1}{8\pi}\frac{\Gamma(4N)}{\Gamma(4(N-1))} E^{-4} \left( 1 - \frac{p_j^0}{E} \right)^{4N-5} \,.
\end{equation}

One can easily chech that this result is correctly normalized, i.e. $\int d^4p_j f(p_j) = 1$. The density of states with energy $p_j^0$ can be computed by integrating in $d^3 p_j$, and the result is
\begin{eqnarray}
f(p_j^0) dp_j^0 &=& dp_j^0 \int d^3p_j f(p_j) \, \Theta(p_j^0 - |\vec{p}_j|)  \nonumber \\
&=& dp_j^0 \frac{1}{6}  \frac{\Gamma(4N)}{\Gamma(4(N-1))} E^{-1} \left( \frac{p_j^0}{E}\right)^3 \left( 1 - \frac{p_j^0}{E} \right)^{4N-5} \,.
\end{eqnarray}

If one considers the system in the rest mass frame, then $E = M$, i.e. the total energy of the system is equal to its mass. In addition, $p_j^0 = \varepsilon_j$ is the energy of particle $j$. Then, we can express this result in the equivalent form
\begin{equation}
f(\varepsilon_j) d\varepsilon_j =  \frac{1}{6} \frac{\Gamma(4N)}{\Gamma(4(N-1))}\left( \frac{\varepsilon_j}{M} \right)^3 \left( 1 - \frac{\varepsilon_j}{M} \right)^{4N-5}  d\left( \frac{\varepsilon_j}{M} \right) \,.
\end{equation}


\begin{thebibliography}{1}
\bibitem{Dyson1} F.J. Dyson, Phys. Rev. 75 (1949) 1736.
\bibitem{Gell-Mann_Low} M. Gell-Mann and F.E. Low, Phys. Rev. 95 (1954) 1300.
\bibitem{Ward} J.C. Ward, Proc. Phys. Soc. (London) A64 (1951) 54.
\bibitem{Mandelbrot} Mandelbrot, B.B. The Fractal Geometry of Nature; WH Freeman: New York, NY, USA, 1983.
\bibitem{West} G. West, {\it Scale, The Universal Laws of Life, Growth, and Death in Organisms, Cities, and Companies}, Penguin Books, 2018 - New York.
\bibitem{Falconer} K. Falconer, {\it Fractals, a short introduction}, Oxford Univ. Press, 2013 - Oxford.
\bibitem{Tsallis1988} C. Tsallis, J. Stat. Phys. 52 (1988) 479.
\bibitem{TsallisBook} M.~Gell-Mann, C.~Tsallis, {\it Nonextensive Entropy: Interdisciplinary Applications}, Oxfor University Press, USA, 2004. 
\bibitem{Politzer1} H. D. Politzer, Phys. Reports 14 (1974) 129.
\bibitem{Politzer2} H. Georgi and H. D. Politzer, Phys. Rev. D9 (1974) 416.
\bibitem{GrossWilczek1} D. Gross and F. Wilczek, Phys. Reports 14 (1974) {\bf incomplete!!!}
\bibitem{GrossWilczek2} D. Gross and F. Wilczek, Phys. Rev. D9 (1974) 980.
\bibitem{Callan} C. G. Callan, Jr.,  Phys. Rev. D 2 (1970) 1541.
\bibitem{Symanzik1} K. Symanzik, Commun. math. Phys. 18 (1970) 227.
\bibitem{Symanzik2} K. Symanzik, Commun. math. Phys. 23 (1971) 49.
\bibitem{DeppmanMegiasPhysics} A. Deppman and E. Megias, Physics 2019, 1, 103-110.
\bibitem{Deppman2016} A. Deppman, Phys. Rev. D 93 (2016) 054001.
\bibitem{DFMM} A. Deppman, T. Frederico, E. Meg\'{\i}as, D. P Menezes, Entropy 20 (2018) 633.  
\bibitem{Hagedorn0} R. Hagedorn, CERN - TH. 520 65/166/5 (1965).
\bibitem{Hagedorn65} R. Hagedorn, Nuovo Cimento Suppl. 3 (1965) 147.
\bibitem{Deppman2012} A. Deppman, Physica A 391 (2012) 6380.
\bibitem{Feal} X. Feal, C. Pajares, C and R.A.  Vazquez, Phys. Rev. C 99 (2019) 015205.
\bibitem{PedroCardoso} 
  P.~H.~G.~Cardoso, T.~Nunes da Silva, A.~Deppman and D.~P.~Menezes, Eur.\ Phys.\ J.\ A 53, no. 10, 191 (2017).
\bibitem{Megias} E. Meg\'{\i}as, D. P. Menezes, and A. Deppman, Physica A 421, 15 (2015).
\bibitem{Debora} D.~P.~Menezes, A.~Deppman, E.~Meg\'{\i}as and L.~B.~Castro, Eur.\ Phys.\ J.\ A 51, no. 12, 155 (2015).
\bibitem{Deppman_Universe} A. Deppman, Universe 3 (2017)  62.

\bibitem{Tokarev} M. Tokarev and I. Zborovsky, EPJ Web Conf. 141 (2017) 02006.
\bibitem{Zborovsky} I. Zborovský and M.V. Tokarev, Phys. Rev. D75, (2007) 094008.
\bibitem{WWselfsymmetry} G. Wilk and Z. W\l{}odarczyk, Phys. Lett. B (2013) 163.

\bibitem{BeckCosmicRay} G. Yalcin and C. Beck, Sci. Rep. 8 (2018) 1764.
\bibitem{Gross:1973id} 
  D.~J.~Gross and F.~Wilczek,
  Phys.\ Rev.\ Lett.\  {\bf 30}, 1343 (1973).
\bibitem{Politzer:1974fr} 
  H.~D.~Politzer,
  Phys.\ Rept.\  {\bf 14}, 129 (1974).
\bibitem{Sena} I. Sena and a. Deppman, Eur. Phys. J. A 49 (2013) 17.
\bibitem{Lucas} L. Marques, E. Andrade-II and A. Deppman, Phys. Rev. D 87 (2013) 114022.
\bibitem{Lucas2} L. Marques, J. Cleymans and A. Deppman, Phys. Rev. D 91 (2015) 054025.
\bibitem{vanRitbergen:1997va} 
  T.~van Ritbergen, J.~A.~M.~Vermaseren and S.~A.~Larin,
  Phys.\ Lett.\ B {\bf 400}, 379 (1997).
\bibitem{Czakon:2004bu} 
  M.~Czakon,
  Nucl.\ Phys.\ B {\bf 710}, 485 (2005).
\bibitem{WilkWlodarczyk} M.~Rybczynski, Z.~Wlodarczyk and G.~Wilk, J.\ Phys.\ G 39, 095004 (2012).
\bibitem{Bialas_Peschanski} A. Bialas and R. Peschanki, Nucl. Phys. B273 (1986) 703.
\bibitem{Bialas_Peschanski2} A. Bialas and R. Peschanki, Nucl. Phys. B308 (1988) 857.
\bibitem{Hwa} R. C. Hwa, Phys. Rev. D 41 (1990) 1456.
\bibitem{HwaPan} R.C. Hwa and J. Pan, Phys. Rev. D 45 (1992) 1476.
\bibitem{Hegyi1} S. Hegyi, Phys. Lett. B 318 (1993) 642.
\bibitem{DreminHwa} I.M. Dremin and R.C. Hwa, Phys. Rev. D 49 (1994) 5805.
\bibitem{Hegyi2} S. Hegyi and T. Cs\"org\"o, Phys. Lett. B 296 (1992) 256.
\bibitem{Antoniou} N.G. Antoniou, N. Davis and F.K. Diakonos, Phys. Rev. C 93 (2016) 014908.
\bibitem{Wolf} E.A. De Wolf, I.M. Dremin, W. Kittel, Phys. Rep. 270 (1996) 1-141.
\bibitem{Kittel} W. Kittel, E.A. De Wolf, Soft Multihadron Dynamics (World Scientific, 2005).
\bibitem{Sarkisyan_Multiplicity} Edward K.G. Sarkisyan, Aditya Nath Mishra, Raghunath Sahoo and Alexander S. Sakharov, Phys. Rev. D 93 (2016) 054046.




%
\end{thebibliography}
\end{document}